\input hyperbasics
\catcode`\@=11
\def\unredoffs{\voffset=13mm \hoffset=6.5truemm} 
\def\redoffs{\voffset=-12.truemm\hoffset=-3truemm} 
\def\speclscape{}
%
\newbox\leftpage \newdimen\fullhsize \newdimen\hstitle \newdimen\hsbody
\newdimen\hdim
\hfuzz=1pt
\ifx\hyperdef\UNd@FiNeD\def\hyperdef#1#2#3#4{#4}\def\hyperref#1#2#3#4{#4}\fi
\def\newans{y }
\def\answb{y }
\ifx\answb\newans\message{(This uses normal fonts.)}%
%
\def\bigans{b }
%
\def\answ{b }
\ifx\answ\bigans\message{(Format simple colonne 12pts.}
\magnification=1200 \unredoffs\hsize=147truemm\vsize=219truemm
\hsbody=\hsize \hstitle=\hsize 
\else\message{(Format double colonne, 10pts.} \let\l@r=L
\magnification=1000 \vsize=182.5truemm
\redoffs%
\hstitle=122.5truemm\hsbody=122.5truemm\fullhsize=258truemm\hsize=\hsbody 
\output={
  \almostshipout{\leftline{\vbox{\makeheadline\pagebody\makefootline}}}
\advancepageno%
}
\def\almostshipout#1{\if L\l@r \count1=1 \message{[\the\count0.\the\count1]}
      \global\setbox\leftpage=#1 \global\let\l@r=R
 \else \count1=2
  \shipout\vbox{\speclscape{\hsize\fullhsize}
      \hbox to\fullhsize{\box\leftpage\hfil#1}}  \global\let\l@r=L\fi}
\fi

\def\sla#1{\mkern-1.5mu\raise0.4pt\hbox{$\not$}\mkern1.2mu #1\mkern 0.7mu}
\def\Dbar{\mkern-1.5mu\raise0.4pt\hbox{$\not$}\mkern-.1mu {\rm D}\mkern.1mu}
\def\Abar{\mkern1.mu\raise0.4pt\hbox{$\not$}\mkern-1.3mu A\mkern.1mu}
\def\Bbar{\mkern-0.mu\raise0.4pt\hbox{$\not$}\mkern-.3mu B\mkern 0.6mu}
\newskip\tableskipamount \tableskipamount=8pt plus 3pt minus 3pt


\newdimen\chapskip

\font\ssbx=cmssbx10  

\font\caprm=cmr9
\font\capit=cmti9
\font\capbf=cmbx9
\font\capsl=cmsl9
\font\capmi=cmmi9
\font\capex=cmex9
\font\capsy=cmsy9
\chapskip=17.5mm
\def\makeheadline{\vbox to 0pt{\vskip-22.5pt
\line{\vbox to8.5pt{}\the\headline}\vss}\nointerlineskip}
\font\tenbi=cmmib10 
\font\ninebi=cmmib9
\font\sevenbi=cmmib7 
\font\fivebi=cmmib5
\textfont4=\tenbi
\scriptfont4=\sevenbi
\scriptscriptfont4=\fivebi
\font\headrm=cmr10

\font\sixrm=cmr6
\font\fiverm=cmr5
\font\sixmi=cmmi6
\font\fivemi=cmmi5
\font\sixsy=cmsy6
\font\fivesy=cmsy5
\font\sixbf=cmbx6
\font\fivebf=cmbx5
\skewchar\capmi='177 \skewchar\sixmi='177 \skewchar\fivemi='177
\skewchar\capsy='60 \skewchar\sixsy='60 \skewchar\fivesy='60

\def\elevenpoint{
\textfont0=\caprm \scriptfont0=\sixrm \scriptscriptfont0=\fiverm
\def\rm{\fam0\caprm}
\textfont1=\capmi \scriptfont1=\sixmi \scriptscriptfont1=\fivemi
\textfont2=\capsy \scriptfont2=\sixsy \scriptscriptfont2=\fivesy
\textfont3=\capex \scriptfont3=\capex \scriptscriptfont3=\capex
\textfont\itfam=\capit \def\it{\fam\itfam\capit} 
\textfont\slfam=\capsl  \def\sl{\fam\slfam\capsl} 
\textfont\bffam=\capbf \scriptfont\bffam=\sixbf
\scriptscriptfont\bffam=\fivebf
\def\bf{\fam\bffam\capbf} 
\textfont4=\ninebi \scriptfont4=\sevenbi \scriptscriptfont4=\fivebi
\abovedisplayskip=11pt plus 3pt minus 8pt
\belowdisplayskip=\abovedisplayskip
\smallskipamount=2.7pt plus 1pt minus 1pt
\medskipamount=5.4pt plus 2pt minus 2pt
\bigskipamount=10.8pt plus 3.6pt minus 3.6pt
\normalbaselineskip=11pt
\setbox\strutbox=\hbox{\vrule height7.8pt depth3.2pt width0pt}
\normalbaselines \rm}

%
%

\catcode`\@=11

\font\tenmsa=msam10
\font\sevenmsa=msam7
\font\fivemsa=msam5
\font\tenmsb=msbm10
\font\sevenmsb=msbm7
\font\fivemsb=msbm5
\newfam\msafam
\newfam\msbfam
\textfont\msafam=\tenmsa  \scriptfont\msafam=\sevenmsa
  \scriptscriptfont\msafam=\fivemsa
\textfont\msbfam=\tenmsb  \scriptfont\msbfam=\sevenmsb
  \scriptscriptfont\msbfam=\fivemsb

\def\hexnumber@#1{\ifcase#1 0\or1\or2\or3\or4\or5\or6\or7\or8\or9\or
	A\or B\or C\or D\or E\or F\fi }

\font\teneuf=eufm10
\font\seveneuf=eufm7
\font\fiveeuf=eufm5
\newfam\euffam
\textfont\euffam=\teneuf
\scriptfont\euffam=\seveneuf
\scriptscriptfont\euffam=\fiveeuf
\def\frak{\ifmmode\let\next\frak@\else
 \def\next{\Err@{Use \string\frak\space only in math mode}}\fi\next}
\def\goth{\ifmmode\let\next\frak@\else
 \def\next{\Err@{Use \string\goth\space only in math mode}}\fi\next}
\def\frak@#1{{\frak@@{#1}}}
\def\frak@@#1{\fam\euffam#1}

\edef\msa@{\hexnumber@\msafam}
\edef\msb@{\hexnumber@\msbfam}

\mathchardef\boxdot="2\msa@00
\mathchardef\boxplus="2\msa@01
\mathchardef\boxtimes="2\msa@02
\mathchardef\square="0\msa@03
\mathchardef\blacksquare="0\msa@04
\mathchardef\centerdot="2\msa@05
\mathchardef\lozenge="0\msa@06
\mathchardef\blacklozenge="0\msa@07
\mathchardef\circlearrowright="3\msa@08
\mathchardef\circlearrowleft="3\msa@09
\mathchardef\rightleftharpoons="3\msa@0A
\mathchardef\leftrightharpoons="3\msa@0B
\mathchardef\boxminus="2\msa@0C
\mathchardef\Vdash="3\msa@0D
\mathchardef\Vvdash="3\msa@0E
\mathchardef\vDash="3\msa@0F
\mathchardef\twoheadrightarrow="3\msa@10
\mathchardef\twoheadleftarrow="3\msa@11
\mathchardef\leftleftarrows="3\msa@12
\mathchardef\rightrightarrows="3\msa@13
\mathchardef\upuparrows="3\msa@14
\mathchardef\downdownarrows="3\msa@15
\mathchardef\upharpoonright="3\msa@16

\mathchardef\downharpoonright="3\msa@17
\mathchardef\upharpoonleft="3\msa@18
\mathchardef\downharpoonleft="3\msa@19
\mathchardef\rightarrowtail="3\msa@1A
\mathchardef\leftarrowtail="3\msa@1B
\mathchardef\leftrightarrows="3\msa@1C
\mathchardef\rightleftarrows="3\msa@1D
\mathchardef\Lsh="3\msa@1E
\mathchardef\Rsh="3\msa@1F
\mathchardef\rightsquigarrow="3\msa@20
\mathchardef\leftrightsquigarrow="3\msa@21
\mathchardef\looparrowleft="3\msa@22
\mathchardef\looparrowright="3\msa@23
\mathchardef\circeq="3\msa@24
\mathchardef\succsim="3\msa@25
\mathchardef\gtrsim="3\msa@26
\mathchardef\gtrapprox="3\msa@27
\mathchardef\multimap="3\msa@28
\mathchardef\therefore="3\msa@29
\mathchardef\because="3\msa@2A
\mathchardef\doteqdot="3\msa@2B

\mathchardef\triangleq="3\msa@2C
\mathchardef\precsim="3\msa@2D
\mathchardef\lesssim="3\msa@2E
\mathchardef\lessapprox="3\msa@2F
\mathchardef\eqslantless="3\msa@30
\mathchardef\eqslantgtr="3\msa@31
\mathchardef\curlyeqprec="3\msa@32
\mathchardef\curlyeqsucc="3\msa@33
\mathchardef\preccurlyeq="3\msa@34
\mathchardef\leqq="3\msa@35
\mathchardef\leqslant="3\msa@36
\mathchardef\lessgtr="3\msa@37
\mathchardef\backprime="0\msa@38
\mathchardef\risingdotseq="3\msa@3A
\mathchardef\fallingdotseq="3\msa@3B
\mathchardef\succcurlyeq="3\msa@3C
\mathchardef\geqq="3\msa@3D
\mathchardef\geqslant="3\msa@3E
\mathchardef\gtrless="3\msa@3F
\mathchardef\sqsubset="3\msa@40
\mathchardef\sqsupset="3\msa@41
\mathchardef\vartriangleright="3\msa@42
\mathchardef\vartriangleleft="3\msa@43
\mathchardef\trianglerighteq="3\msa@44
\mathchardef\trianglelefteq="3\msa@45
\mathchardef\bigstar="0\msa@46
\mathchardef\between="3\msa@47
\mathchardef\blacktriangledown="0\msa@48
\mathchardef\blacktriangleright="3\msa@49
\mathchardef\blacktriangleleft="3\msa@4A
\mathchardef\vartriangle="0\msa@4D
\mathchardef\blacktriangle="0\msa@4E
\mathchardef\triangledown="0\msa@4F
\mathchardef\eqcirc="3\msa@50
\mathchardef\lesseqgtr="3\msa@51
\mathchardef\gtreqless="3\msa@52
\mathchardef\lesseqqgtr="3\msa@53
\mathchardef\gtreqqless="3\msa@54
\mathchardef\Rrightarrow="3\msa@56
\mathchardef\Lleftarrow="3\msa@57
\mathchardef\veebar="2\msa@59
\mathchardef\barwedge="2\msa@5A
\mathchardef\doublebarwedge="2\msa@5B
\mathchardef\angle="0\msa@5C
\mathchardef\measuredangle="0\msa@5D
\mathchardef\sphericalangle="0\msa@5E
\mathchardef\varpropto="3\msa@5F
\mathchardef\smallsmile="3\msa@60
\mathchardef\smallfrown="3\msa@61
\mathchardef\Subset="3\msa@62
\mathchardef\Supset="3\msa@63
\mathchardef\Cup="2\msa@64

\mathchardef\Cap="2\msa@65

\mathchardef\curlywedge="2\msa@66
\mathchardef\curlyvee="2\msa@67
\mathchardef\leftthreetimes="2\msa@68
\mathchardef\rightthreetimes="2\msa@69
\mathchardef\subseteqq="3\msa@6A
\mathchardef\supseteqq="3\msa@6B
\mathchardef\bumpeq="3\msa@6C
\mathchardef\Bumpeq="3\msa@6D
\mathchardef\lll="3\msa@6E

\mathchardef\ggg="3\msa@6F

\mathchardef\circledS="0\msa@73
\mathchardef\pitchfork="3\msa@74
\mathchardef\dotplus="2\msa@75
\mathchardef\backsim="3\msa@76
\mathchardef\backsimeq="3\msa@77
\mathchardef\complement="0\msa@7B
\mathchardef\intercal="2\msa@7C
\mathchardef\circledcirc="2\msa@7D
\mathchardef\circledast="2\msa@7E
\mathchardef\circleddash="2\msa@7F
\def\ulcorner{\delimiter"4\msa@70\msa@70 }
\def\urcorner{\delimiter"5\msa@71\msa@71 }
\def\llcorner{\delimiter"4\msa@78\msa@78 }
\def\lrcorner{\delimiter"5\msa@79\msa@79 }
\def\yen{\mathhexbox\msa@55 }
\def\checkmark{\mathhexbox\msa@58 }
\def\circledR{\mathhexbox\msa@72 }
\def\maltese{\mathhexbox\msa@7A }
\mathchardef\lvertneqq="3\msb@00
\mathchardef\gvertneqq="3\msb@01
\mathchardef\nleq="3\msb@02
\mathchardef\ngeq="3\msb@03
\mathchardef\nless="3\msb@04
\mathchardef\ngtr="3\msb@05
\mathchardef\nprec="3\msb@06
\mathchardef\nsucc="3\msb@07
\mathchardef\lneqq="3\msb@08
\mathchardef\gneqq="3\msb@09
\mathchardef\nleqslant="3\msb@0A
\mathchardef\ngeqslant="3\msb@0B
\mathchardef\lneq="3\msb@0C
\mathchardef\gneq="3\msb@0D
\mathchardef\npreceq="3\msb@0E
\mathchardef\nsucceq="3\msb@0F
\mathchardef\precnsim="3\msb@10
\mathchardef\succnsim="3\msb@11
\mathchardef\lnsim="3\msb@12
\mathchardef\gnsim="3\msb@13
\mathchardef\nleqq="3\msb@14
\mathchardef\ngeqq="3\msb@15
\mathchardef\precneqq="3\msb@16
\mathchardef\succneqq="3\msb@17
\mathchardef\precnapprox="3\msb@18
\mathchardef\succnapprox="3\msb@19
\mathchardef\lnapprox="3\msb@1A
\mathchardef\gnapprox="3\msb@1B
\mathchardef\nsim="3\msb@1C
\mathchardef\ncong="3\msb@1D

\mathchardef\varsubsetneq="3\msb@20
\mathchardef\varsupsetneq="3\msb@21
\mathchardef\nsubseteqq="3\msb@22
\mathchardef\nsupseteqq="3\msb@23
\mathchardef\subsetneqq="3\msb@24
\mathchardef\supsetneqq="3\msb@25
\mathchardef\varsubsetneqq="3\msb@26
\mathchardef\varsupsetneqq="3\msb@27
\mathchardef\subsetneq="3\msb@28
\mathchardef\supsetneq="3\msb@29
\mathchardef\nsubseteq="3\msb@2A
\mathchardef\nsupseteq="3\msb@2B
\mathchardef\nparallel="3\msb@2C
\mathchardef\nmid="3\msb@2D
\mathchardef\nshortmid="3\msb@2E
\mathchardef\nshortparallel="3\msb@2F
\mathchardef\nvdash="3\msb@30
\mathchardef\nVdash="3\msb@31
\mathchardef\nvDash="3\msb@32
\mathchardef\nVDash="3\msb@33
\mathchardef\ntrianglerighteq="3\msb@34
\mathchardef\ntrianglelefteq="3\msb@35
\mathchardef\ntriangleleft="3\msb@36
\mathchardef\ntriangleright="3\msb@37
\mathchardef\nleftarrow="3\msb@38
\mathchardef\nrightarrow="3\msb@39
\mathchardef\nLeftarrow="3\msb@3A
\mathchardef\nRightarrow="3\msb@3B
\mathchardef\nLeftrightarrow="3\msb@3C
\mathchardef\nleftrightarrow="3\msb@3D
\mathchardef\divideontimes="2\msb@3E
\mathchardef\varnothing="0\msb@3F
\mathchardef\nexists="0\msb@40
\mathchardef\mho="0\msb@66
\mathchardef\eth="0\msb@67
\mathchardef\eqsim="3\msb@68
\mathchardef\beth="0\msb@69
\mathchardef\gimel="0\msb@6A
\mathchardef\daleth="0\msb@6B
\mathchardef\lessdot="3\msb@6C
\mathchardef\gtrdot="3\msb@6D
\mathchardef\ltimes="2\msb@6E
\mathchardef\rtimes="2\msb@6F
\mathchardef\shortmid="3\msb@70
\mathchardef\shortparallel="3\msb@71
\mathchardef\smallsetminus="2\msb@72
\mathchardef\thicksim="3\msb@73
\mathchardef\thickapprox="3\msb@74
\mathchardef\approxeq="3\msb@75
\mathchardef\succapprox="3\msb@76
\mathchardef\precapprox="3\msb@77
\mathchardef\curvearrowleft="3\msb@78
\mathchardef\curvearrowright="3\msb@79
\mathchardef\digamma="0\msb@7A
\mathchardef\varkappa="0\msb@7B
\mathchardef\hslash="0\msb@7D
\mathchardef\hbar="0\msb@7E
\mathchardef\backepsilon="3\msb@7F
\def\Bbb{\ifmmode\let\next\Bbb@\else
 \def\next{\errmessage{Use \string\Bbb\space only in math mode}}\fi\next}
\def\Bbb@#1{{\Bbb@@{#1}}}
\def\Bbb@@#1{\fam\msbfam#1}

\catcode`\@=12

\def\sla#1{\mkern-1.5mu\raise0.4pt\hbox{$\not$}\mkern1.2mu #1\mkern 0.7mu}
\def\Dbar{\mkern-1.5mu\raise0.4pt\hbox{$\not$}\mkern-.1mu {\rm D}\mkern.1mu}
\def\Abar{\mkern1.mu\raise0.4pt\hbox{$\not$}\mkern-1.3mu A\mkern.1mu}
\nopagenumbers
\headline={\ifnum\pageno=1\hfill\else\draftdate\hfil{\headrm\folio}%
\hfil\fi}	 
\else\message{(This uses pseudo 12pts fonts.}
\hoffset=8mm
\voffset=16mm
\input lfont12 

\def\sla#1{\mkern-1.5mu\raise0.5pt\hbox{$\not$}\mkern1.2mu #1\mkern 0.7mu}
\def\Dbar{\mkern-1.5mu\raise0.5pt\hbox{$\not$}\mkern-.1mu {\rm D}\mkern.1mu}
\def\Abar{\mkern1.mu\raise0.5pt\hbox{$\not$}\mkern-1.3mu A\mkern.1mu}
\fi

\newcount\yearltd\yearltd=\year\advance\yearltd by -1900
\newif\ifdraftmode
\draftmodefalse
\def\draft{\draftmodetrue{\count255=\time\divide\count255 by 60
\xdef\hourmin{\number\count255} 
  \multiply\count255 by-60\advance\count255 by\time
  \xdef\hourmin{\hourmin:\ifnum\count255<10 0\fi\the\count255}}}
\def\draftdate{\ifdraftmode{\headrm\quad (\jobname,\ le
\number\day/\number\month/\number\yearltd\ \ \hourmin)}\else{}\fi} 
\newif\iffrancmode
\francmodefalse
\def\e{\mathop{\rm e}\nolimits}

\def\d{{\rm d}}

\def\tr{\mathop{\rm tr}\nolimits}
\def\det{\mathop{\rm det}\nolimits}

\chardef\sigmat=27
\def\n{\noindent}  
\def\rf{\par\item{}}
\def\nrf{\par\n}
\def\frac#1#2{{\textstyle{#1\over#2}}}

\def\leaderfill{\leaders\hbox to 1em{\hss.\hss}\hfill}
\catcode`\@=11
\def\deqalignno#1{\displ@y\tabskip\centering \halign to
\displaywidth{\hfil$\displaystyle{##}$\tabskip0pt&$\displaystyle{{}##}$
\hfil\tabskip0pt &\quad
\hfil$\displaystyle{##}$\tabskip0pt&$\displaystyle{{}##}$ 
\hfil\tabskip\centering& \llap{$##$}\tabskip0pt \crcr #1 \crcr}}
\def\deqalign#1{\null\,\vcenter{\openup\jot\m@th\ialign{
\strut\hfil$\displaystyle{##}$&$\displaystyle{{}##}$\hfil
&&\quad\strut\hfil$\displaystyle{##}$&$\displaystyle{{}##}$
\hfil\crcr#1\crcr}}\,}
\def\xlabel#1{\expandafter\xl@bel#1}\def\xl@bel#1{#1}
\def\label#1{\l@bel #1{\hbox{}}}
\def\l@bel#1{\ifx\UNd@FiNeD#1\message{label \string#1 is undefined.}%
\xdef#1{??? }\fi{\let\hyperref=\relax\xdef\next{#1}}%
\ifx\next\em@rk\def\next{}%
\else\def\next{#1}\fi\next}
\def\DefWarn#1{\ifx\UNd@FiNeD#1\else
\immediate\write16{*** WARNING: the label \string#1 is already defined%
***}\fi}%
\newread\ch@ckfile
\def\cinput#1{\def\filen@me{#1 }
\immediate\openin\ch@ckfile=\filen@me
\ifeof\ch@ckfile\message{<< (\filen@me\ DOES NOT EXIST in this pass)>>}\else%
\closein \ch@ckfile\input\filen@me\fi}
\newread\ch@ckfile
\immediate\openin\ch@ckfile=\jobname.def
\ifeof\ch@ckfile\message{<< (\jobname.def DOES NOT EXIST in this pass) >>}
\else
\def\DefWarn#1{}%
\closein \ch@ckfile%
\input\jobname.def\fi
\def\listcontent{
\immediate\openin\ch@ckfile=\jobname.tab 
\ifeof\ch@ckfile\message{no file \jobname.tab, no table of contents this
pass}%
\else\closein\ch@ckfile\centerline{\bf\iffrancmode Table des
mati\`eres \else Contents\fi}\nobreak\medskip%
{\baselineskip=12pt\parskip=0pt\catcode`\@=11\input\jobname.tab
\catcode`\@=12\bigbreak\bigskip}\fi}
\newcount\nosection
\newcount\nosubsection
\newcount\neqno
\newcount\notenumber
\newcount\nofigure
\newif\ifappmode
\def\equation{\jobname.equ}
\newwrite\equa

\newdimen\hulp
\def\maketitle#1{
\edef\oneliner##1{\centerline{##1}}
\edef\twoliner##1{\vbox{\parindent=0pt\leftskip=0pt plus 1fill\rightskip=0pt
plus 1fill 
                     \parfillskip=0pt\relax##1}} 
\setbox0=\vbox{#1}\hulp=0.5\hsize
                 \ifdim\wd0<\hulp\oneliner{#1}\else
                 \twoliner{#1}\fi}
\def\preprint#1{\ifdraftmode\gdef\prepname{\jobname/#1}\else%
\gdef\prepname{#1}\fi\hfill{
\expandafter{\prepname}}\vskip20mm} 
\def\title#1\par{\gdef\titlename{#1}
\maketitle{\ssbx\uppercase\expandafter{\titlename}}
\vskip20truemm
\nosection=0
\neqno=0
\notenumber=0
\nofigure=0
\def\prefix{}
\appmodefalse
\mark{\the\nosection}
\message{#1}
\immediate\openout\equa=\equation
}
\def\abstract{\vskip8mm\iffrancmode\centerline{RESUME}\else%
\centerline{ABSTRACT}\fi \smallskip \begingroup\narrower
\elevenpoint\baselineskip10pt} 
\def\endabstract{\par\endgroup \bigskip}
\def\section#1\par{\vskip0pt plus.1\vsize\penalty-100\vskip0pt plus-.1
\vsize\bigskip\vskip\parskip
\ifnum\nosection=0\ifappmode\relax\else\writetoc
\fi\fi
\advance\nosection by 1\global\nosubsection=0\global\neqno=0
\vbox{\noindent\bf{\hyperdef\hypernoname{section}{\prefix\the\nosection}%
{\prefix\the\nosection}\ #1}}
\writetoca{{\string\hyperref{}{section}{\prefix\the\nosection}%
{\prefix\the\nosection}} {#1}}
\message{\the\nosection\ #1}
\mark{\the\nosection}\bigskip\noindent
}

\def\appendix#1#2\par{\bigbreak\nosection=0
\appmodetrue
\notenumber=0
\neqno=0
\def\prefix{A}
\mark{\the\nosection}
\message{APPENDICES}
{\leftline{APPENDICES} \hyperdef\hypernoname{appendix}{app}{ 
\leftline{\uppercase\expandafter{#1}}
\leftline{\uppercase\expandafter{#2}}}}
\bigskip\noindent\nonfrenchspacing
\writetoca{\string\hyperref{}{appendix}{app}{Appendices}.\ #1.\ #2}%
}
\def\subsection#1\par {\vskip0pt plus.05\vsize\penalty-100\vskip0pt
plus-.05\vsize\bigskip\vskip\parskip\advance\nosubsection by 1
\vbox{\noindent\it{\hyperdef\hypernoname{subsection}{\prefix\the\nosection.%
\the\nosubsection}{\prefix\the\nosection.\the\nosubsection\ #1}}}%
\smallskip\noindent 
\writetoca{{\string\hyperref{}{subsection}{\prefix\the\nosection.%
\the\nosubsection}{\prefix\the\nosection.\the\nosubsection}} {#1}}
\message{\the\nosection.\the\nosubsection\ #1}
} 
\def\note #1{\advance\notenumber by 1
\footnote{$^{\the\notenumber}$}{\sevenrm #1}} 

\parindent=1em 
\newinsert\margin
\dimen\margin=\maxdimen
\count\margin=0 \skip\margin=0pt
\def\sslbl#1{\DefWarn#1%
\ifdraftmode{\hfill\escapechar-1{\rlap{\hskip-1mm%
\sevenrm\string#1}}}\fi%
\ifnum\nosection=0\if\prefix{}\xdef#1{}%
\edef\ewrite{\write\equa{{\string#1}}%
\write\equa{}}\ewrite%
\else
\xdef#1{\noexpand\hyperref{}{section}{\prefix}{\prefix}}%
\edef\ewrite{\write\equa{{\string#1},\prefix}%
\write\equa{}}\ewrite%
\writedef{#1\leftbracket#1}
\fi
\else%
\ifnum\nosubsection=0%
\xdef#1{\noexpand\hyperref{}{section}{\prefix\the\nosection}{\prefix\the\nosection}}%
\edef\ewrite{\write\equa{{\string#1},\prefix\the\nosection}%
\write\equa{}}\ewrite%
\writedef{#1\leftbracket#1}
\else%
\xdef#1{\noexpand\hyperref{}{subsection}{\prefix\the\nosection.%
\the\nosubsection}{\prefix\the\nosection.\the\nosubsection}}%
\writedef{#1\leftbracket#1}
\edef\ewrite{\write\equa{{\string#1},\prefix\the\nosection%
.\the\nosubsection}\write\equa{}}\ewrite\fi\fi}%

\newwrite\tfile \def\writetoca#1{}
\def\writetoc{\immediate\openout\tfile=\jobname.tab
   \def\writetoca##1{{\edef\next{\write\tfile{\noindent ##1
   \string\leaderfill {\string\hyperref{}{page}{\noexpand\number\pageno}%
                       {\noexpand\number\pageno}} \par}}\next}}}

%
\def\nolabels{\def\wrlabeL##1{}\def\eqlabeL##1{}\def\reflabeL##1{}}
\def\writelabels{\def\wrlabeL##1{\leavevmode\vadjust{\rlap{\smash%
{\line{{\escapechar=` \hfill\rlap{\sevenrm\hskip.03in\string##1}}}}}}}%
\def\eqlabeL##1{{\escapechar-1\rlap{\sevenrm\hskip.05in\string##1}}}%
\def\reflabeL##1{\noexpand\llap{\noexpand\sevenrm\string\string\string##1}}}
\nolabels

\global\newcount\refno \global\refno=1
\newwrite\rfile
\def\ref{[\hyperref{}{reference}{\the\refno}{\the\refno}]\nref}
\def\nref#1{\DefWarn#1%
\xdef#1{[\noexpand\hyperref{}{reference}{\the\refno}{\the\refno}]}%
\writedef{#1\leftbracket#1}%
\ifnum\refno=1\immediate\openout\rfile=\jobname.ref\fi
\chardef\wfile=\rfile\immediate\write\rfile{\noexpand\item{[\noexpand\hyperdef%
\noexpand\hypernoname{reference}{\the\refno}{\the\refno}]\ }%
\reflabeL{#1\hskip.31in}\pctsign}\global\advance\refno by1\findarg}
\def\findarg#1#{\begingroup\obeylines\newlinechar=`\^^M\pass@rg}
{\obeylines\gdef\pass@rg#1{\writ@line\relax #1^^M\hbox{}^^M}%
\gdef\writ@line#1^^M{\expandafter\toks0\expandafter{\striprel@x #1}%
\edef\next{\the\toks0}\ifx\next\em@rk\let\next=\endgroup\else\ifx\next\empty%
\else\immediate\write\wfile{\the\toks0}\fi\let\next=\writ@line\fi\next\relax}}
\def\striprel@x#1{} \def\em@rk{\hbox{}}
\def\lref{\begingroup\obeylines\lr@f}
\def\lr@f#1#2{\DefWarn#1\gdef#1{\let#1=\UNd@FiNeD\ref#1{#2}}\endgroup\unskip}

\def\addref#1{\immediate\write\rfile{\noexpand\item{}#1}} 
\def\listrefs{{}\vfill\supereject\immediate\closeout\rfile\writestoppt
\baselineskip=14pt\centerline{{\bf\iffrancmode R\'eferences\else References%
\fi}}
\bigskip{\parindent=20pt%
\frenchspacing\escapechar=` \input \jobname.ref\vfill\eject}\nonfrenchspacing}
\def\startrefs#1{\immediate\openout\rfile=\jobname.ref\refno=#1}
\def\xref{\expandafter\xr@f}\def\xr@f[#1]{#1}
\def\refs#1{\count255=1[\r@fs #1{\hbox{}}]}
\def\r@fs#1{\ifx\UNd@FiNeD#1\message{reflabel \string#1 is undefined.}%
\nref#1{need to supply reference \string#1.}\fi%
\vphantom{\hphantom{#1}}{\let\hyperref=\relax\xdef\next{#1}}%
\ifx\next\em@rk\def\next{}%
\else\ifx\next#1\ifodd\count255\relax\xref#1\count255=0\fi%
\else#1\count255=1\fi\let\next=\r@fs\fi\next}
%
\newwrite\lfile
{\escapechar-1\xdef\pctsign{\string\%}\xdef\leftbracket{\string\{}
\xdef\rightbracket{\string\}}\xdef\numbersign{\string\#}}
\def\writedefs{\immediate\openout\lfile=\jobname.def \def\writedef##1{%
{\let\hyperref=\relax\let\hyperdef=\relax\let\hypernoname=\relax
 \immediate\write\lfile{\string\def\string##1\rightbracket}}}}%
\def\writestop{\def\writestoppt{\immediate\write\lfile{\string\pageno%
\the\pageno\string\startrefs\leftbracket\the\refno\rightbracket%
\string\def\string\secsym\leftbracket\secsym\rightbracket%
\string\secno\the\secno\string\meqno\the\meqno}\immediate\closeout\lfile}}
\def\writestoppt{}\def\writedef#1{}
\writedefs
\def\biblio\par{\vskip0pt plus.1\vsize\penalty-100\vskip0pt plus-.1
\vsize\bigskip\vskip\parskip
\message{Bibliographie}
{\leftline{\bf \hyperdef\hypernoname{biblio}{bib}{Bibliographical Notes}}}
\nobreak\medskip\noindent\frenchspacing
\writetoca{\string\hyperref{}{biblio}{bib}{Bibliographical Notes}}}%

\def\biblionote{\iffrancmode Notes Bibliographiques\else Bibliographical Notes
\fi}
\def\beginbib\par{\vskip0pt plus.1\vsize\penalty-100\vskip0pt plus-.1
\vsize\bigskip\vskip\parskip
\message{Bibliographie}
{\leftline{\bf \hyperdef\hypernoname{biblio}{\the\nosection}%
{\biblionote}}}
\nobreak\medskip\noindent\frenchspacing
\writetoca{\string\hyperref{}{biblio}{\the\nosection}%
{\biblionote}}}%
\def\endbib{\nonfrenchspacing}

\def\Exercises{\iffrancmode Exercices\else Exercises
\fi}
\def\exerc\par{\vskip0pt plus.1\vsize\penalty-100\vskip0pt plus-.1
\vsize\bigskip\vskip\parskip
\message{Exercises}
{\leftline{\bf \hyperdef\hypernoname{exercise}{\the\nosection}{\Exercises}}}
\nobreak\medskip\noindent\frenchspacing
\writetoca{\string\hyperref{}{exercise}{\the\nosection}{\Exercises}}
}
\def\eqnn{\global\advance\neqno by 1 \ifinner\relax\else%
\eqno\fi(\prefix\the\nosection.\the\neqno)}
%
\def\eqnd#1{\DefWarn#1%
\global\advance\neqno by 1 
{\xdef#1{($\noexpand\hyperref{}{equation}{\prefix\the\nosection.\the\neqno}%
{\prefix\the\nosection.\the\neqno}$)}}
\ifinner\relax\else\eqno\fi(\hyperdef\hypernoname{equation}{\prefix\the%
\nosection.\the\neqno}{\prefix\the\nosection.\the\neqno})
\writedef{#1\leftbracket#1}
\ifdraftmode{\escapechar-1{\rlap{\hskip.2mm\sevenrm\string#1}}}\fi
\edef\ewrite{\write\equa{{\string#1},(\prefix\the\nosection.\the\neqno)
{\noexpand\number\pageno}}\write\equa{}}\ewrite}
%
\def\checkm@de#1#2{\ifmmode{\def\f@rst##1{##1}\hyperdef\hypernoname{equation}%
{#1}{#2}}\else\hyperref{}{equation}{#1}{#2}\fi}
\def\f@rst#1{\c@t#1a\em@ark}\def\c@t#1#2\em@ark{#1}
\def\eqna#1{\DefWarn#1%
\global\advance\neqno by1\ifdraftmode{\hfill%
\escapechar-1{\rlap{\sevenrm\string#1}}}\fi%
\xdef #1##1{(\noexpand\relax\noexpand%
\checkm@de{\prefix\the\nosection.\the\neqno\noexpand\f@rst{##1}1}%
{\hbox{$\prefix\the\nosection.\the\neqno##1$}})}
\writedef{#1\numbersign1\leftbracket#1{\numbersign1}}%
} 
%

%
\def\em@rk{\hbox{}} 
\def\xeqn{\expandafter\xe@n}\def\xe@n(#1){#1}
\def\xeqna#1{\expandafter\xe@na#1}\def\xe@na\hbox#1{\xe@nap #1}
\def\xe@nap$(#1)${\hbox{$#1$}}
\def\eqns#1{(\e@ns #1{\hbox{}})}
\def\e@ns#1{\ifx\UNd@FiNeD#1\message{eqnlabel \string#1 is undefined.}%
\xdef#1{(?.?)}\fi{\let\hyperref=\relax\xdef\next{#1}}%
\ifx\next\em@rk\def\next{}%
\else\ifx\next#1\xeqn#1\else\def\n@xt{#1}\ifx\n@xt\next#1\else\xeqna#1\fi
\fi\let\next=\e@ns\fi\next}
\def\figure#1#2{\global\advance\nofigure by 1 \vglue#1%
{\elevenpoint
\setbox1=\hbox{#2}
\ifdim\wd1=0pt\centerline{Fig.\ \the\nofigure\hskip0.5mm}%
\else\def\caption{Fig.\ \the\nofigure\quad#2\hskip0mm}
\setbox0=\hbox{\caption}
\ifdim\wd0>\hsize\noindent Fig.\ \the\nofigure\quad#2\else
                 \centerline{\caption}\fi\fi}\par}
\def\lfigure#1#2{\global\advance\nofigure by
1\vglue#1\leftline{\elevenpoint\hskip10truemm  Fig.\
\the\nofigure\quad #2}} 
\def\figlbl#1{\ifdraftmode{\hfill\escapechar-1{\rlap{\hskip-1mm%
\sevenrm\string#1}}}\fi%
{\xdef#1{\noexpand\hyperref{}{figure}{\the\nofigure}%
{\the\nofigure}}}%
\edef\ewrite{\write\equa{{\string#1}\the\nofigure}%
\write\equa{}}\ewrite%
\writedef{#1\leftbracket#1}}

\catcode`@=12


\def\phib{\phi}
\def\lambdab{\lambda}
\preprint{T99/058}

\title{Renormalization of gauge theories and master equation}

\centerline{J.~ZINN-JUSTIN*}
\medskip{\it
CEA-Saclay, Service de Physique Th\'eorique**, F-91191 Gif-sur-Yvette
\goodbreak Cedex, FRANCE} 

\footnote{}{${}^*$email: zinn@spht.saclay.cea.fr}

\footnote{}{${}^{**}$Laboratoire de la Direction des
Sciences de la Mati\`ere du 
Commissariat \`a l'Energie Atomique}

\abstract
The evolution of ideas which has led from the first proofs of the
renormalizability of non-abelian gauge theories, based on Slavnov--Taylor
identities, to the modern proof based on the BRS symmetry and the {\it master
equation}\/ is recalled. This lecture has been delivered at the {\bf Symposium
in the Honour of Professor C.~N.~Yang}, Stony-Brook, May 21-22 1999.
\endabstract
\vfill\eject
\section Introduction

It is a rare privilege for me to open this conference in honour of Professor
Yang. His scientific contributions have been for me an essential source of
inspiration. The most obvious example, Yang--Mills fields or gauge theories,
will be illustrated by my talk. But there are other important aspects of
Pr.~Yang's work which have also directly influenced me: Professor Yang has
consistently shown us that 
a theorist could contribute to quite different domains of physics like
Particle Physics, the Statistical Physics of phase transitions or integrable
systems.... Moreover his work has always emphasized mathematical elegance.
  
Finally by offering me a position at the ITP in Stony-Brook in 1971, Pr.~Yang
has given me the opportunity to start with the late Benjamin W.~Lee a work
on the renormalization of gauge theories, which has kept me busy for
several years and played a major role in my scientific career. 
        
Let me add a few other personal words. The academic year 1971--1972 I
spent here at the ITP has been of the most exciting and memorable of
my scientific life. One reason of course is my successful
collaboration with Ben Lee. However another reason is the specially
stimulating atmosphere Professor Yang had managed to create at the ITP, by
attracting talented physicists, both ITP members and visitors, by the
style of scientific discussions, seminars and lectures.

My interest in Yang--Mills fields actually dates back to 1969, and in
1970 I started a work, very much in the spirit of the original paper
of Yang and Mills, on the application  of massive Yang-Mills fields to
Strong Interaction dynamics. Although in our work massive Yang-Mills
fields were treated in the spirit of effective field theories, we were aware
of the fact that such a quantum field theory was not renormalizable. 

In the summer of 1970 I presented the preliminary results of 
our work in a summer school in Carg\`ese, where Ben Lee was lecturing on
the renormalization of spontaneous and linear symmetry breaking. This
had the consequence that one year later I arrived here at the ITP to
work with him. 

Ben had just learned, in a conference I believe, from 't Hooft's
latest work on the renormalizability of non-abelian gauge theories
both in the symmetric and spontaneously broken phase and was busy
proving renormalizability of the abelian Higgs model. We immediately
started our work on the much more involved non-abelian extension. 

Our work was based on functional integrals and functional
methods and a generalization of so-called Slavnov--Taylor identities,
consequence of the properties of the Faddeev--Popov (FP) determinant
arising in the quantization of gauge theories. In a series of four
papers (1972--1973), we examined most aspects of the renormalization of
gauge theories. 
\section Classical gauge action and quantization

The principle of gauge invariance which promotes
continuous global (or rigid) symmetries to local (gauge) symmetries provides a
beautiful geometrical method to generate interactions between particles. The
pure Yang--Mills action has the form
$${\cal S} \left({\bf A}_{\mu} \right) = -{1\over 4e^{2}} \int \d^{d}x\, \tr
{\bf F}^2_{\mu \nu} \left(x \right) ,  $$ 
where ${\bf A}_\mu(x)$ is the gauge field, a matrix belonging to the Lie
algebra of the symmetry group, and ${\bf F}_{\mu\nu}(x)$ the associated
curvature obtained from the covariant derivative ${\bf D}_\mu$ 
$${\bf D}_\mu=\partial_\mu +{\bf A}_\mu\,, $$
by
$$ {\bf F}_{\mu\nu}(x) = \left[ {\bf D}_{\mu},{\bf
D}_{\nu}\right] = \partial_{\mu}{\bf  A}_{\nu} -  \partial_{\nu}{\bf
A}_{\mu}  + \left[ {\bf A}_{\mu},{\bf A}_{\nu}\right]. $$
Matter fields which transform non-trivially under the group will then be
coupled to the gauge field. For fermions the action takes the typical form
$$ {\cal S}_{\rm F} \left(\bar \psi ,\psi \right) = - \int \d  ^{d}x\, \bar
\psi \left(x \right) \left(\sla{{\bf D}} + M \right) 
\psi \left(x \right), $$
and for the boson fields:
$$  {\cal S}_{\rm B}(\phib) = \int \d  ^{d}x \left[ \left(
{\bf D}_{\mu}\phib\right)^{\dagger}{\bf D} _{\mu}\phib + V \left( \phib
\right) \right],  $$ 
in which $V(\phib)$ is a group invariant function of the scalar field $\phib$.
\medskip
{\it Quantization.} The classical action results from a beautiful
construction, but the quantization apparently completely destroys the
geometric structure. Due to the gauge
invariance the degrees of freedom associated with gauge transformations have
no dynamics and therefore a straightforward quantization of the classical
action does not generate a meaningful perturbation theory (though
non-perturbative calculations in lattice regularized gauge theories can be
performed). 
It is thus necessary to {\it fix}\/ the gauge, a way of expressing that some
dynamics has to be provided for these degrees of freedom. For example,
motivated by Quantum Electrodynamics, one may add to the action a covariant
non-gauge invariant contribution 
$${\cal S}_{\rm gauge}={1 \over 2\xi e^2}\int\d^d x\,\tr
\left(\partial_{\mu}{\bf A}_{\mu}\right)^2. \eqnd\enaLanga $$
However, simultaneously, and this is a specificity of non-abelian gauge
theories, it is necessary to modify the functional integration measure
of the gauge field to maintain formal unitarity. In the case of Landau's gauge
\enaLanga~one finds 
$$[\d{\bf A}_\mu(x)]\mapsto [\d{\bf A}_\mu(x)]\det{\bf M}\,, \eqnd\eFPdet $$
where $\bf M$ is the operator
$${\bf M}(x,y)=\partial_\mu{\bf D}_\mu\delta(x-y).$$
This (Faddeev--Popov) determinant is the source of many difficulties. Indeed
after quantizing the theory one has to renormalize it. Renormalization is a
theory of deformations of local actions. However the determinant generates
a non-local contribution to the action. Of course, using a well-known trick,
it is possible to rewrite the determinant as resulting from the integration
over un-physical spin-less fermions ${\bf C},\bar{\bf C}$ (the ``ghosts")
of an additional contribution to the action
$${\cal S}_{\rm ghosts}=\int\d^d x\,\bar{\bf C}(x)\partial_\mu{\bf D}_\mu
{\bf C}(x).$$
After this transformation the action is local and renormalizable in the sense
of power counting. However, in this local form all traces of the original
symmetry seem to have been lost.
\section Renormalization

The measure \eFPdet~is the invariant measure for a set of non-local
transformations which for infinitesimal transformations takes the form
$$\delta {\bf A}_\mu(x)=\int \d y\,{\bf D}_\mu {\bf M}^{-1}(x,y)\omega(y),$$
the field $\omega(x)$ parametrizing the transformation. Using this property it
is possible to derive a set of Ward--Takahashi (Slavnov--Taylor) identities
between Green's functions and to prove renormalizability of gauge theories
both in the symmetric and spontaneously broken Higgs phase. The non-local
character of these transformations and the necessity of using two different
representations, one non-local but with invariance properties, the other one
local and thus suitable for power counting analysis, explains the complexity
of the initial proofs.  

Though the problem of renormalizing gauge theories could then be considered as
settled, one of the remaining problems was that the proofs, even in
the most synthetic presentation like in Lee--Zinn-Justin IV, were complicated,
non-transparent, and more based on trial and error than systematic
methods.  

Returning to Saclay I tried to systematize the renormalization program
of quantum field theories with symmetries. I abandoned the determination of
renormalization constants by relation between Green's functions, for a more
systematic approach based on loop expansion and counter-terms. 

The idea is to proceed by induction on the number of loops. Quickly 
summarized: 

One starts from a regularized local lagrangian with some symmetry
properties. One derives, as consequence of the symmetry,
identities (generally called Ward--Takahashi or WT identities) satisfied by
the generating functional $\Gamma$ of one-particle irreducible (1PI)
Green's functions (or proper vertices). By letting the cut-off go
to infinity (or the dimension to four in dimensional regularization)
one obtains identities satisfied by the sum $\Gamma_{\rm div}$ of all
divergent contributions at one loop order. At this order
$\Gamma_{\rm div}$ is a local functional of a degree determined by
power counting. By subtracting $\Gamma_{\rm div}$  from the action one
obtains a theory finite at one-loop order. One then reads off the
symmetry of the lagrangian renormalized at one-loop order 
and repeats the procedure to renormalize at two-loop order. The
renormalization program is then based on determining general
identities valid both for the action and the 1PI functional, which
are stable under renormalization, i.e.~stable under all deformations
allowed by power counting. One finally proves the stability by induction
on the number of loops.

Unfortunately this program did not apply to non-abelian gauge
theories, because it required a symmetry of the local quantized action, and
none was apparent. WT identities were established using symmetry properties 
of the theory in the non-local representation
 
In the spring of 1974 my student Zuber drew my attention
to a preliminary report of a work of Becchi, Rouet and Stora who had
discovered a strange fermion-type (like supersymmetry)
symmetry of the complete quantized action including the ghost
contributions. There were indications that this symmetry could be used to
somewhat simplify the algebra of the proof of renormalization. Some time later,
facing the daunting prospect of lecturing about renormalization of gauge
theories and explaining the proofs to non-experts, I decided to study the BRS
symmetry. I then realized that the BRS symmetry was the key allowing the
application of the general renormalization scheme and in a summer school in
Bonn (1974) I presented a general proof of renormalizability of gauge theories
based on BRS symmetry and the {\it master equation}. 
\section BRS symmetry

The form of the BRS transformations in the case of non-abelian gauge
transformations is rather involved and hides its simple origin. We thus give
here a presentation which shows how BRS symmetry arises in apparently a
simpler context. 
Let $ \varphi^\alpha $ be a set of dynamical variables satisfying a system
of equations:
$$ E_\alpha (\varphi)=0\,,\eqnd\eequ  $$
where the functions $E_\alpha (\varphi)$ are smooth, and
$E_\alpha=E_\alpha (\varphi)$ is a one-to-one map in some
neighbourhood of $E_\alpha=0$ which can be inverted in
$\varphi^\alpha=\varphi^\alpha(E)$. This implies in particular that 
the equation \eequ~has a unique solution $ \varphi_{\rm s}^\alpha $. We then
consider some function $ F(\varphi) $ and we look for a formal representation
of $ F \left(\varphi_{\rm s} \right)$, which does not require solving equation
\eequ~explicitly. We can then  write:
$$\eqalignno{ F\left(\varphi_{\rm s} \right)&=\int\biggl\{ \prod_{\alpha}
\d E^\alpha  \, \delta \left( E_\alpha \right) \biggr\}
F\bigl(\varphi(E)\bigr)\cr
&=\int\biggl\{ \prod_{\alpha}
\d \varphi^\alpha  \, \delta \left[ E_\alpha(\varphi) \right] \biggr\}
{\cal J}(\varphi)\, F(\varphi), &\eqnd\erep    \cr} $$  
with:
$${\cal J}(\varphi)=\det{\bf E}\,,\quad E_{\alpha\beta}\equiv{\partial
E_\alpha \over \partial \varphi^\beta}.  $$
We have chosen $ E_\alpha(\varphi) $ such that $ \det{\bf E}$ is positive. 
\medskip
{\it Slavnov--Taylor identity}. The measure
$ \d \rho (\varphi)$:
$$ \d\rho(\varphi)={\cal J}(\varphi) \prod_{\alpha} \d\varphi^\alpha \,,
\eqnd{\edet} $$ 
has a simple property. The measure $\prod_\alpha\d
E_\alpha $ is the invariant measure for the group of translations
$E_\alpha \mapsto E_\alpha+\nu_\alpha$.
It follows that $ \d \rho (\varphi)$ is the invariant measure for the
translation  
group non-linearly realized on the new coordinates $\varphi_\alpha$
(provided $\nu_\alpha$ is small enough):
$$ \varphi^\alpha  \mapsto \varphi'{}^\alpha \quad
{\rm with} \quad E_\alpha\left(\varphi'\right)-\nu_\alpha =E_\alpha 
(\varphi). \eqnd{\eslt} $$ 
This is the origin, in gauge theories, of the Slavnov--Taylor symmetry. \par 
The infinitesimal form of the transformation law can be written more
explicitly: 
$$ \delta \varphi^\alpha = [E^{-1}(\varphi)]^{\alpha\beta}
\nu_{\beta}\,. \eqnd\estinf $$
\medskip
{\it BRS symmetry.}
Let us again start from identity \erep\ and first replace the $
\delta $-function by its Fourier representation:
$$ \prod_{\alpha}\delta \left[E_\alpha (\varphi)\right] = \int
\prod_{\alpha}{\d\lambda^\alpha  \over 2i\pi} \e^{-\lambda^{\alpha}
E_\alpha(\varphi)}.  \eqnn $$
The $ \lambda $-integration runs along the imaginary axis. From the rules of
fermion integration we know that we can also write the determinant as an
integral over Grassmann variables $ c^\alpha  $ and $ \bar c^{\alpha}$:
$$ \det{\bf E}= \int \prod_{\alpha} \left(\d c^\alpha \d\bar c^{\alpha}
\right)\exp\left(\bar c^{\alpha}E_{\alpha\beta} c^{\beta}\right). \eqnn $$
Expression \erep~then takes the apparently more complicated form 
$$ F\left(\varphi_{\rm s}\right) ={\cal N} \int \prod_{\alpha}
\left(\d \varphi^\alpha\d c^\alpha \d \bar c^{\alpha} \d
\lambda^{\alpha} \right)F(\varphi) \exp\left[-S (\varphi, c,\bar
c,\lambda)\right], \eqnd{\erepb} $$
in which $\cal N $ is a constant normalization factor and $ S
\left(\varphi,c,\bar c ,\lambda \right) $ the quantity:  
$$ S\left(\varphi ,c,\bar c ,\lambda \right)=\lambda^{\alpha}E_\alpha 
(\varphi)-\bar c^{\alpha}E_{\alpha\beta}(\varphi) c^\beta\,.
\eqnd{\eactef} $$
While we seem to have replaced a simple problem by a more complicated one, in
fact in many situations (and this includes the case where equation \eequ~is a 
field equation) it is easy to work with the integral representation \erepb.\par
Quite surprisingly the function $S$ has a symmetry, which actually
is a consequence of the invariance of the measure \edet~under the group of
transformations \estinf. This BRS symmetry, first discovered in the
quantization of gauge theories by Becchi, Rouet and Stora (BRS), is a
fermionic symmetry in the sense that it transforms commuting variables into 
Grassmann variables and vice versa. The parameter of the transformation is a
Grassmann variable, an anti-commuting constant $ \bar \varepsilon $. The
variations of the various dynamic variables are:
$$\left\lbrace\vcenter{\openup1\jot\halign{$\hfil#$&${}#\hfil$&\qquad
$\hfil#$&${}#\hfil$ 
\cr \delta\varphi^\alpha =&\bar\varepsilon c^\alpha \, ,&
\delta c^\alpha =& 0\,, \cr 
\delta \bar c^\alpha =& \bar \varepsilon\lambda^\alpha\, , &\delta
\lambda^\alpha = & 0\,, \cr}} \right. \eqnd{\etrbrs} $$
with:
$$ \bar \varepsilon^{2}=0\, ,\qquad \bar \varepsilon c^\alpha +
c^\alpha \bar \varepsilon =0 \, ,\qquad \bar \varepsilon\bar c^{\alpha}+\bar
\varepsilon\bar c^{\alpha}=0\,. $$ 
The transformation is obviously {\it nilpotent}\/ of vanishing square:
$\delta^2=0$. \par  
The BRS transformation can be represented by a Grassmann differential
operator ${\cal D}$, when acting on functions of $\{\varphi,c,\bar
c,\lambda\}$: 
$${\cal D}=c^\alpha {\partial \over \partial \varphi^\alpha }+
\lambda^{\alpha}{\partial \over \partial\bar c^{\alpha}}\, . \eqnd{\opbrs} $$ 
The nil-potency of the BRS transformation is then expressed by the identity:
$${\cal D}^2=0\,. \eqnd{\enilpott} $$
\section The master equation
  
In gauge theories the role of the $\varphi$ variables is played by the
group elements which parametrize gauge transformations and the equation
\eequ~is simply the gauge fixing equation. The form of the BRS transformation
is more complicated only because it is written in terms of group elements:
$$\left\lbrace\vcenter{\openup1\jot\halign{$\hfil#$&${}#\hfil$&\qquad

$\hfil#$&${}#\hfil$ \cr\delta {\bf A}_{\mu}(x)   &= -\bar\varepsilon {\bf
D}_{\mu}{\bf C}(x)\,, & \delta{\bf C}(x) & =\bar \varepsilon{\bf C}
^{2}(x),\cr 
\delta \bar{\bf C}(x)& =\bar \varepsilon \lambdab(x) , &  \delta \lambdab(x)
 &=0\,. \cr} } \right. \eqnd{\eBRSgr} $$
However this form of BRS transformations is not stable under
renormalization because the form of the gauge transformations is modified by
the renormalization. \par
To discuss renormalization it is necessary to add to the action two sources
${\bf K}_\mu$, $\bf L$, for the BRS transformations which are not linear in the
fields 
$${\cal S}\mapsto {\cal S}+\int\d^4 x\tr \left(-{\bf K}_\mu(x) {\bf
D}_{\mu}{\bf C}(x)+{\bf L}(x){\bf C}^2(x)\right). $$
The sources for BRS transformations, ${\bf K}_\mu$ and $\bf L$, have been
later renamed anti-fields.\par
The stable relation satisfied by the complete action, including these
additional contributions, then takes a, at first sight disappointingly simple,
quadratic form (here written in a simple example, without matter fields)    
$$\int d^4 x\left( {\delta {\cal S}\over \delta A_\mu^\alpha (x) }{\delta {\cal
S}\over \delta K_\mu^\alpha (x) }  +{\delta {\cal S}\over \delta
C^\alpha(x)}{\delta {\cal S} \over \delta L^\alpha(x)}+\lambda^\alpha(x)
{\delta {\cal S}\over \delta\bar C^\alpha(x)} \right)=0\,.\eqnd\eYMmaster $$
 In particular
the master equation \eYMmaster~contains no explicit reference to the initial
gauge transformation. Therefore one may worry that it does not determine
the renormalized action completely, and that the general renormalization
program fails in the case of non-abelian gauge theories. However, one slowly
discovers that the master equation has remarkable properties. In particular
all its local solutions which satisfy the power counting requirements, have
indeed the form of an action for a quantized non-abelian gauge theory. Then
continuity implies, in the semi-simple example at least, preservation of all
geometric properties.  
  
One surprising outcome still bothered me for some time: The master
equation has solutions with quartic ghost interactions, which cannot be
obviously related to a determinant. On the other hand the master
equation by itself (and this one of its main properties) implies gauge
independence and unitarity.  

Only a few years later, elaborating on a remark of Slavnov, was I able to
reproduce a general quartic ghost term as resulting from a generalized
gauge fixing procedure (Zinn-Justin 1984). 

After the renormalization program was successfully completed, one important
problem remained, of relevance for instance to the description
of deep-inelastic scattering experiments: the renormalization
of gauge invariant operators of dimension higher than four.   
Using similar techniques Stern-Kluberg and Zuber were able to solve
the problem for operators of dimension six and conjecture the general
form. Only recently has the general conjecture been proven rigourously by
non-trivial cohomology techniques (Barnich, Brandt and Henneaux 1995). 
\bigskip
{\bf Acknowledgments.} To Professor C.N.~Yang as a testimony of admiration and
gratitude, Stony-Brook, May 21, 1999. 
\bigskip
\beginbib

\nrf After the fundamental article \rf
C.N.~Yang and R.L.~Mills, {\it Phys.~Rev.} 96 (1954) 191, 
\nrf the main issue was the quantization of gauge theories\rf
R.P. Feynman, {\it Acta Phys. Polon.} 24 (1963) 697;\par
B.S. DeWitt, {\it Phys. Rev.} 162 (1967) 1195, 1239;\par
L.D. Faddeev and V.N. Popov, {\it Phys. Lett.} 25B (1967) 29;\par
S. Mandelstam, {\it Phys. Rev.} 175 (1968) 1580.\par
\nrf In the following article we tried to apply the idea of massive
Yang--Mills fields to Strong Interaction Dynamics\rf
J.L.~Basdevant and J.~Zinn-Justin, {\it Phys.~Rev.} D3 (1971) 1865.
\nrf Among the articles discussing Ward--Takahashi and renormalization see for
instance\rf 
G. 't Hooft, {\it Nucl. Phys.} B33 (1971) 173; {\it ibid.} B35 (1971) 167.\par
A.A. Slavnov, {\it Theor. Math. Phys.} 10 (1972) 99.\par
J.C. Taylor, {\it Nucl. Phys.} B33 (1971) 436.\par
B.W. Lee and J. Zinn-Justin, {\it Phys. Rev.} D5 (1972) 3121, {\it ibid.}
3137,  {\it ibid.} 3155;  {\it ibid.} D7 (1973) 1049.\par
G. 't Hooft and M. Veltman, {\it Nucl. Phys.} B50 (1972) 318.\par
D.A. Ross and J.C. Taylor,  {\it Nucl. Phys.} B51 (1973) 125;\par
B.W. Lee, {\it Phys. Lett.} 46B (1973) 214; {\it Phys. Rev.} D9 (1974) 933.
\nrf The anti-commuting type symmetry of the quantized action is exhibited in
\rf
C. Becchi, A. Rouet and R. Stora, {\it Comm. Math. Phys.} 42 (1975) 127.
\nrf Most of the preceding articles are reprinted in \rf
Selected papers on {\it Gauge Theory of Weak and Electromagnetic
Interactions}, C.H.~Lai ed., World Scientific (Singapore 1981).
\nrf The general proof, based on BRS symmetry and the master equation, of
renormalizability in an arbitrary gauge, can be found in the
proceedings of the Bonn summer school 1974,\rf
J. Zinn-Justin in {\it Trends in Elementary Particle Physics (Lectures Notes
in Physics} 37),  H. Rollnik and K. Dietz eds. (Springer-Verlag,
Berlin 1975).
\nrf See also\rf
J.~Zinn-Justin in {\it Proc. of the 12th School of Theoretical Physics,
Karpacz 1975}, Acta Universitatis   Wratislaviensis 368; \par
B.W. Lee in {\it Methods in Field Theory}, Les Houches 1975, R. Balian and J.
Zinn-Justin eds. (North-Holland, Amsterdam 1976).
\nrf Finally a systematic presentation can be found in\rf
J.~Zinn-Justin, {\it Quantum Field Theory and Critical Phenomena},
Clarendon Press (Oxford 1989, third ed.~1996). 
\nrf For an an alternative proof based on BRS symmetry and the BPHZ formalism
see \rf
C. Becchi, A. Rouet and R. Stora, {\it Ann. Phys. (NY)} 98 (1976) 287. 
\nrf Non-linear gauges and the origin of quartic ghost terms are investigated
in \rf 
J. Zinn-Justin, {\it Nucl. Phys.} B246 (1984) 246.
\nrf Renormalization of gauge invariant operators and the BRST cohomology are
discussed in\rf
H.~Kluberg-Stern and J.-B. Zuber, {\it Phys. Rev.} D12 (1975) 467;\par
G.~Barnich, and M.~Henneaux, {\it Phys. Rev. Lett.} 72 (1994) 1588;
G.~Barnich, F.~Brandt and M.~Henneaux, {\it Comm. Math. Phys.} 174 (1995) 93. 
\endbib
\bye